# Second Harmonic Generation in Lithiated Silicon Nanowires: Derivations and Computational Methods

**Donald C Boone***

***- Nanoscience Research Institute -** Alexandria, Virginia 22314 USA. Email: db2585@caa.columbia.edu

**Abstract:** This research will examine the computational methods to calculate the nonlinear optical process of second harmonic generation (SHG) that will be hypothesized to be present during lithium ion insertion into silicon nanowires. First it will be determined whether the medium in which SHG is conveyed is non-centrosymmetric or whether the medium is inversion symmetric where SHG as a part of the second-order nonlinear optical phenomenon does not exist. It will be demonstrated that the main interaction that determines SHG is multiphoton absorption on lithium ions. The quantum harmonic oscillator (QHO) is used as the background that generates coherent states for electrons and photons that transverse the length of the silicon nanowire. The matrix elements of the Hamiltonian which represents the energy of the system will be used to calculate the probability density of second-order nonlinear optical interactions which includes collectively SHG, sum-frequency generation (SFG) and difference-frequency generation (DFG). As a result it will be seen that at varies concentrations of lithium ions (Li+) within the crystallized silicon (c-Si) matrix the second-order nonlinear optical process has probabilities substantial enough to create second harmonic generation that could possibly be used for such applications as second harmonic imaging microscopy.

## Introduction

For over the last decade lithiated silicon nanowires has been extensively researched as being potentially the next great advancement in anode material in lithium ion batteries (LIBs). The increased specific charge capacity from approximately 400 mA-hr/g in carbon based batteries to over 4000 mA-hr/g in LIBs has been documented in many research studies of lithiated processes of silicon nanowires [1,2]. However, the volume increase of over 300% for these nanowires due to lithium ion insertion has caused material fatigue and fractures of these nanowires and ultimate failures. A possible alternative for lithiated silicon nanowires other than energy storage devices are applications with second harmonic generation (SHG). It has been well known for over the last two decades that crystalline silicon (c-Si) can experience anisotropic expansion due to polarization from an electric field [3]. More recently it has been shown that certain crystals exposed to lasers has produced anisotropic expansion in certain crystallized lattices that leads to the creation of second harmonic generation [4]. It has been hypothesized that when an applied electron flux travels through a lithiated silicon nanowire that the ensuing anisotropic volume expansion is caused by spontaneous and stimulated emission which are the same physical processes that are cause by a laser [5].

This computational analysis will begin with an energy source of electrons that will generate an average electromagnetic energy of 2-eV applied to one end of the silicon nanowire. This average energy source will define the structure of the quantum harmonics oscillator (QHO) which serve as our mathematical model. The QHO will be constructed of a series of two electron-volts energy states. On the opposite end of the silicon nanowire lithium ions will be inserted to simulate the lithiated diffusion process.

The electrons that moves through the silicon lattice generates a quantized electromagnetic field represented by the manifestation of photons. When multiple photons are absorbed by a lithium ion, the ion experiences an excitation that transitions the lithium ion from the ground state to an excited state. Once the lithium ion transitions to an elevated energy state, it is subjected to the spontaneous emission process [6].

The continuation of photons that absorbs into lithium ions causing there excitation and at the same time the diffusion process of these ions causes an increase in the lithiated silicon density. The total atomic system in our lithiated silicon lattice model experiences population inversion which is define as a majority of atoms or ions being in the excited state. When the lithium ions are in such a state with photons being transmitted and absorbed within this dense lithium-silicon particle matrix, populated inversion becomes the prelude to the stimulated emission process [7]. Stimulated emission occurs when an incoming photon interact with a lithium ion in the excited state inducing it to transition an electron to the ground state emitting a photon that is approximately of the same angular frequency, phase and direction of the incoming photon. Since the lithiated silicon nanowire is modeled after a quantum harmonic oscillator, these photons are said to be in a coherent state with angular frequency ω. These photons represents the electromagnetic mode and are analogous to oscillating waves [8].

During second harmonic generation (SHG) the two photons of the stimulated emission process $\omega_1$ and $\omega_2$ has angular frequencies that are equal $\omega_1$=$\omega_2$. These two photons interfere constructively during their interaction resulting in photon $\omega_3$ where $\omega_3 = \omega_1 + \omega_2 = 2\omega$ as energy is conserved. The phase matching condition for the corresponding wave vectors of the photons is $\overrightarrow{k_3} = \overrightarrow{k_1} + \overrightarrow{k_2}$ in which there momentum is also conserved [9].

It will be displayed in the following research that the solution of the Schrödinger equation when it is solved as an eigenvalue system of equations will lead to a group of probability density states within nonlinear optical interactions. One such group will be the *second-order* probability density states which is the focus of these computational methods that will lead to second harmonic generation.

As the lithium diffusion process continues, the accumulation of lithium ions within the silicon nanowire defines the lithium ion concentration *x* which is the ratio of the number of lithium ions to silicon atoms per unit volume. It will be demonstrated in this research that if the lithium ion concentration is maintain at a level $x$ < 2.00 then second harmonic generation can be developed.

## Non-Centrosymmetric Medium

The prerequisite for establishing all second-order nonlinear optical interactions which includes SHG is to determine if the medium in which it is manifested in is non-centrosymmetric which are materials that do not conform to inversion symmetry. Materials that are identified as centrosymmetric are fluids, most amorphous solids and many crystals and are therefore thought of as inversion symmetric [10]. The crystalline silicon (c-Si) that composes the nanowire and lithium ions (Li+) that are modeled after a fluid are both separately centrosymmetric materials and therefore has the property of inversion symmetry. However, once lithium ions are diffused through the crystal silicon lattice structure, the lithium ions 'breaks the symmetry' of the c-Si and these two materials combined to form a non-centrosymmetric medium and thus supportive of second-order nonlinear optical processes such as SHG. In order to determine the state of the material, an equation has been derived to calculate the inversion symmetry of the medium. This calculation is based on probability theory where the material symmetry is model after a normal distributed Gaussian function. Since the wave function is also a Gaussian function, an equivalence between the two can be established. The wave function for the silicon and lithium particles are constructed by Slater determinant and can be stated as

$$\Psi_A(r,\theta,\phi) = \frac{1}{\sqrt{N_A!}}\left(\sum_{1}^{M}(-1)^{n+m}B\prod_{1}^{N} r^{(n-1)}e^{-Z_{\mathrm{eff}}\left(\frac{r}{a_o n}\right)}Y(\theta,\phi)\right) \quad (1)$$

The wave function $\Psi_A$ is written in this form in order to display that in essence the wave function is a Gaussian equation. The subscript A denotes the species of the particle (c-Si, Li+ or amorphous lithiated silicon- a-$Li_x$Si), $r, \theta, \phi$ are the spherical coordinates, *N* is the number of energy states, *M* is the highest order of the nonlinear optical process, $Z_{\mathrm{eff}}$ is the effective atomic number, B is the normalization constant, $N_A$ is the number of ions and/or atoms in the system, $a_o$ is Bohr radius and $Y(\theta, \phi)$ is the spherical harmonics. The wave function written as a normal distribution function is stated as

$$\Psi_A = C_A \prod_{1}^{N_A} e^{-\frac{1}{2}\left(\frac{r-\mu}{\sigma}\right)^2} \qquad (2)$$

where $C_A$ is the wave function coefficient which ensure that $\Psi_A$ is a solution to the Schrödinger equation and $(r_{ij} - \mu/\sigma)$ is defined as the standardized normal variate (SNV) or variate for brevity [11]. This leads to the statistical parameter that is a measure of symmetry called skewness $\mu_3$which is the expectation value of the cube of the variate derived from equations (1) and (2)

$$\mu_3 = \left|\left\langle \Psi_A \middle| \left[-\frac{1}{2N_A} ln\left(\left|\frac{\Psi_A}{C_A}\right|\right)\right]^{\frac{3}{2}} \middle| \Psi_A \right\rangle\right| \qquad (3)$$

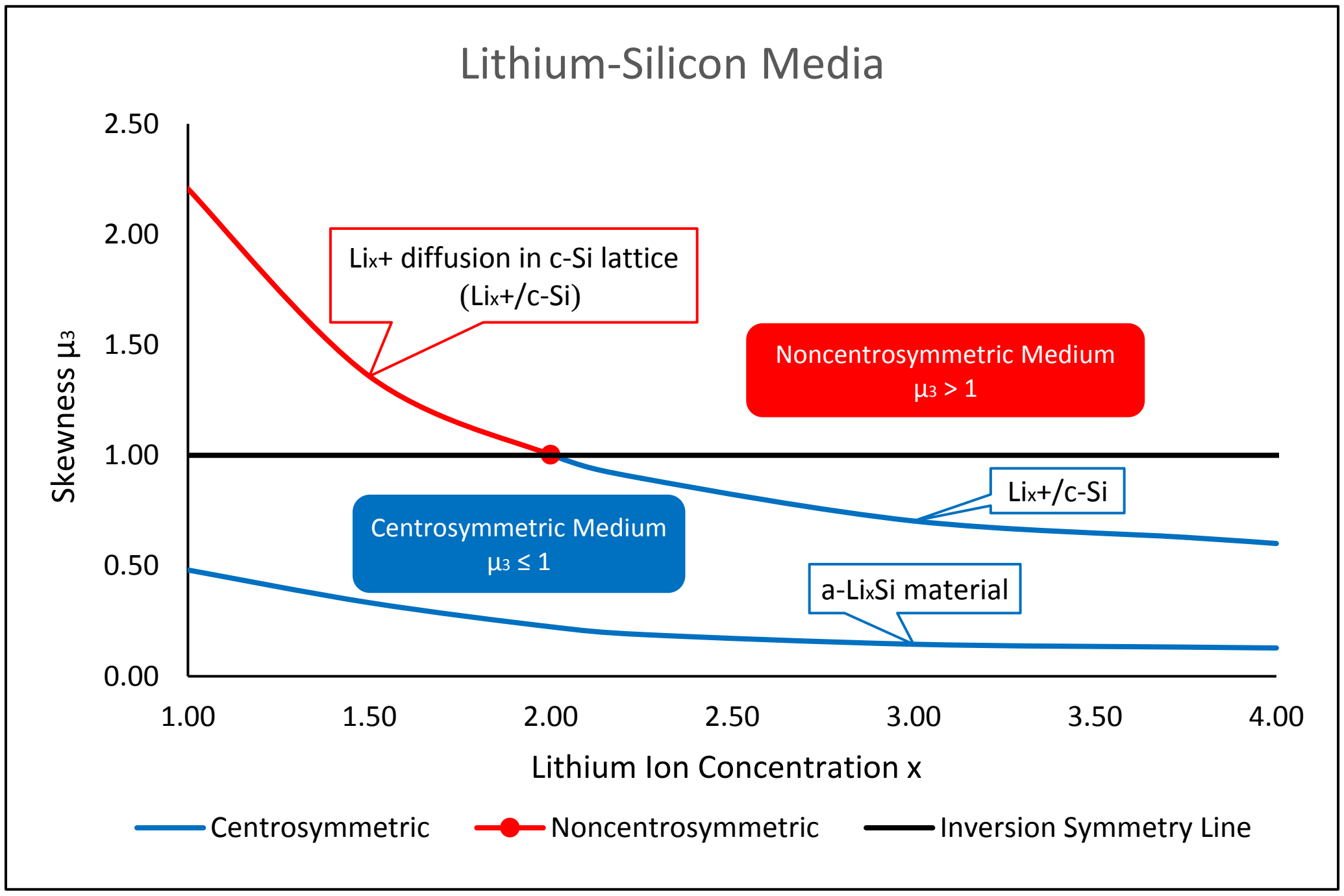

**Figure 1.** During the lithiation process of silicon the symmetry of lithium-silicon medium is measured by the dimensionless parameter skewness $\mu_3$. The centrosymmetric medium has a skewness of $\mu_3 \leq 1$ which is the molecular amorphous lithium-silicon material a-$Li_x$Si. The non-centrosymmetric medium is when the lithium ions Li+ diffuses through the crystallizes silicon c-Si lattice ($Li_x$+/c-Si) and is represented by a skewness of $\mu_3 > 1$. The red dot at lithium ion concentration $x$=2.00 represents the transition point of the lithium-silicon ($Li_x$+/c-Si) media from non-centrosymmetric to centrosymmetric medium.

Historically skewness $\mu_3$ has been called the third standardized moment in probability theory [11]. The wave function $\Psi_A$ for each individual material whether it's atoms, ions or molecules can be calculated by a variety of methods including density functional theory (DFT) however for this study the wave functions were formulated using Hartree-Fock approximation using MATLAB® version R2020a mathematical software.

The existence of second-order nonlinear optical phenomena can be found in any medium with a skewness that is greater than one. This is displayed in figure 1 with the diffusion of lithium ions through the silicon lattice medium ($Li_x$+/c-Si). It is seen that as the lithium ion concentration $x$ increases the skewness $\mu_3$ decreases. At a point where $x$=2.00, the skewness $\mu_3 = 1.00$. A further increase in lithium ion concentration would decrease $\mu_3$ below one and therefore the ($Li_x$+/c-Si) medium will become centrosymmetric and unable to support second-order nonlinear processes. Also shown in figure 1 is amorphous lithiated silicon (a-$Li_xSi$) which is centrosymmetric since its skewness is less than one. The difference between ($Li_x$+/c-Si) and a-$Li_xSi$ is that the former is lithium ions Li+ diffuses through c-Si and are not bonded to any silicon atoms and the latter are complex random molecules that are composed of lithium ions and silicon atoms that are either formed by ionic or covalent bonding. The crystalline silicon c-Si prior to lithiation when the concentration is $x$=0 has a skewness less than 1.0 and is constant at $\mu_3 = 0.689$. Only when lithium ions are introduced to the silicon nanowire is when c-Si is no longer inversion symmetric with the condition that the concentration of Li+ does not exceed $x$=2.00.

## Polarization

After non-centrosymmetric medium has been established, the energy of the system will be defined by developing the Hamiltonian in order to calculate the probability densities states of nonlinear optical processes. Each matrix element in the Hamiltonian is defined by an energy based on the polarization which is defined by the Taylor series

$$P(t) = \epsilon_o \chi^{(1)} \vec{E}(t) + \epsilon_o \chi^{(2)} \vec{E}^2(t) + \epsilon_o \chi^{(3)} \vec{E}^3(t) + \cdots \quad (4)$$

where P(t) is the polarization, $\epsilon_o$ the electric permittivity, $\chi^{(m)}$ is the electric susceptibility to the $m$-th order and $\vec{E}(t)$ is the electric field. The dimensions of each parameter are of standard metric system with the reminder that for $\chi^{(1)}$ is dimensionless i.e. unity and , $\chi^{(2)}$, $\chi^{(3)}$, $\chi^{(4)}$etc. are of the $m$-th order inverse electric field units i.e. $1/(Volts/mass)$, $1/(Volts/mass)^2$, $1/(Volts/mass)^3$ respectively. The polarization is also defined in terms of electric dipole moment per unit volume

$$P(\omega) = -N_P e(x^{(1)} + x^{(2)} + x^{(3)} + \cdots) \qquad (5)$$

where $N_P$ is the number of electric particles per unit volume, $e$ is the electric charge unit and $x^{(m)}$ is the distance between opposite electric charges – positive lithium ions and negative electrons to the $m$-th order. A system of nonlinear inhomogeneous differential equations are used to solve for $x^{(m)}$ that comes from the work of Boyd [10]. The first three nonlinear orders for the solution of $x^{(m)}$ are displayed below:

$$\frac{d^2x^{(1)}}{dt^2} + 2\gamma \frac{dx^{(1)}}{dt} + D(\omega)x^{(1)} = -\frac{e\vec{E}(t)}{m_{eff}} \qquad (6)$$

$$\frac{d^2x^{(2)}}{dt^2} + 2\gamma \frac{dx^{(2)}}{dt} + D(\omega)x^{(2)} + \frac{\omega_0}{a}\left[x^{(1)}\right]^2 = 0 \qquad (7)$$

$$\frac{d^2x^{(3)}}{dt^2} + 2\gamma \frac{dx^{(3)}}{dt} + D(\omega)x^{(3)} + 2\frac{\omega_0}{a}x^{(1)}x^{(2)} = 0 \qquad (8)$$

The $m$ order of any nonlinear optical interaction will define the number of photons $m$ that one lithium ion will absorb *simultaneously* that will result in excitation of lithium ion and electron in the ground state to a conduction band in the quantum harmonic oscillator. The equation of motion for $x^{(m)}$ when $m$=1 is

$$\frac{d^2x^{(m)}}{dt^2} + 2\gamma \frac{dx^{(m)}}{dt} + D(\omega)x^{(m)} = -\frac{e\vec{E}(t)}{m_{eff}} \qquad (9a)$$

and for m ≥ 2 is

$$\frac{d^2x^{(m)}}{dt^2} + 2\gamma\frac{dx^{(m)}}{dt} + D(\omega)x^{(m)} + (m-1)\frac{\omega_0}{a}x^{(1)}x^{(m-1)} = 0 \qquad (9b)$$

where

$$\vec{E}(t) = \mathrm{i}\mathrm{C_E}\frac{\bar{\mathrm{h}}^2(3\pi^2\bar{n}_c)^{\frac{2}{3}}\mathrm{v}_{DOS}}{4\mathrm{n_v}\mathrm{e}\mathrm{m_{eff}}}[\mathrm{u_c}\nabla\mathrm{u_c^*} - \mathrm{u_c^*}\nabla\mathrm{u_c}]\mathrm{e}^{\frac{\gamma r - i\omega_e t}{2}} \qquad (10)$$

$$D(\omega) = \omega_0^2 - \omega^2 - 2i\omega\gamma \qquad (11)$$

$$\omega_0 = \left(\frac{e^2\bar{n}_c}{3m_{eff}\varepsilon_r}\right)^{\frac{1}{2}} \qquad (12)$$

$$\gamma \approx \tau_U^{-1} = 2\gamma_G^2\frac{k_BT}{\mu V_0}\frac{{\omega_e}^2}{\omega_D} \qquad (13)$$

$$V_0 = \frac{a^3}{N_{Li}+N_{Si}} = \frac{1}{n_{Li}+n_{Si}} \qquad (14)$$

$$\gamma_G = -\frac{a}{3\omega}\frac{\partial\omega_e}{\partial a} \qquad (15)$$

$$\omega_e = \frac{\bar{h}(3\pi^2\bar{N}_C\,a^{-3})^{2/3}}{2m_{eff}} \qquad (16)$$

$$\omega_D = (6\pi^2 v_s n_{Li})^{\frac{1}{3}} \qquad (17)$$

$$v_s = \left(\gamma_{ab}\frac{\bar{N}_C m_{eff} v_d^2}{m_{Li}}\right)^{1/2} \qquad (18)$$

Equation 10 is the electric field $\vec{E}(t)$ which was defined previously in research study Boone [7]. Equation 11 is known as the denominator function $D(\omega)$ where ω is the photon angular frequency of the electric field. In equation 12, $\omega_0$ defines the natural frequency of an electron as it travels through the silicon nanowire, $m_{eff}$ is the effective mass of the electron, the average negative charge differential per unit volume $\bar{n}_c$ is define as the average net charge difference between negatively charge electrons and positively charge lithium ions. The relative electric permittivity $\varepsilon_{rij}$ is the constitutive property that defines how the dielectric material affects an applied electric field. An assumption is made in equation 13 that the dipole damping coefficient $\gamma$ is equivalent to the inverse of the relaxation time $\tau_U$ for Umklapp scattering process (U-process) which is known as a phonon-phonon scattering interaction. This assumption is made due to the anharmonic effects that is experienced by both $\gamma$ and $\tau_U$. For the damping coefficient $\gamma$ the dipole moments that is composed of electrons and lithium ions experiences nonlinear interactions or anharmonic effects due to second-order through eighth-order terms in the polarization that is defined in equation 4 and 5. Equation 13 is also define by Boltzmann constant $k_B$, temperature T of the medium, shear modulus $\mu$, and original volume $V_O$ composing of number of silicon atoms $N_{Si}$ and lithium ions $N_{Li}$ as shown in equation 14 where $a$ is the silicon lattice constant. In addition the relaxation time $\tau_U$ of the U-process is a function of the Gruneisen parameter $\gamma_G$ (equation 15) which is geometry dependent and can be described in the same manner as a Taylor series with anharmonic effect terms similar to the polarization [13]. Since the polarization is defined as the dipole moment per unit volume, both dipole damping coefficient $\gamma$ and the inverse of relaxation time $\tau_U$ can be thought of as being analogues to each other.

Equation 16 defines the average angular frequency $\omega_e$ of an electron where $\overline{N}_C$ is the average negative charge differential, $\bar{h}$ is the Planck constant and the silicon lattice constant is $a$. In equation 17 the Debye frequency $\omega_D$ is defined as a function of the lithium ion density $n_{Li}$ and the speed of sound $v_s$ through c-Si. The parameter $v_s$ in equation 18 is a function of the electron drift velocity $v_d$, lithium ion mass $m_{Li}$ and the adiabatic index $\gamma_{ab}$ which is equal to one since the heat capacity of constant pressure and constant volume are equal $(C_p = C_v)$.

This leads to the solution of $x^{(m)}$ in equations 9a and 9b

$$x^{(m)} = -(m-1)!\left[\frac{e^m \vec{E}(t)^m \omega_0^{(2m-2)}}{a^{(m-1)} m_{eff}^m D^{(2m-1)}}\right] \qquad (19)$$

**Matrix Elements**

Since $x^{(m)}$ defines the distance in the electric dipole moment, the individual matrix elements can be constructed in the Hamiltonian. The energy $E_{nm}$ within each matrix element is divided into two energy components – lithium ions that are responsible for the energy states $E_{nm}^{Li}$within the quantum harmonic oscillator and silicon atoms that composes the energy band structure $E_{nm}^{Si}$ within the crystallized silicon lattice. The components of the energy are defined as

$$E_{nm}^{Li} = S_n^{Li} \vec{E} e \langle u_c | x^{(m)} | u_c \rangle \qquad (20)$$

$$E_{nm}^{Si} = S_n^{Si} \left[\frac{1}{E_g{}^2} \frac{\bar{h}^2}{m_{eff}^2} |\langle u_c | k_{Si} \cdot p | \Psi_{Si} \rangle|^2\right]^{\left(\frac{m+1}{2}\right)} \qquad (21)$$

where

$$S_n^{Si} = 10^{-\left(\log_{10} E_{nl}^{Li} + \frac{3}{2} l\right)} \qquad l = 1,2,3, \ldots .8 \qquad (22)$$

$$\vec{E} = \mathrm{i}\mathrm{C_E} \frac{\bar{\mathrm{h}}^2 (3\pi^2 \bar{n}_c)^{\frac{2}{3}} \mathrm{v}_{DOS}}{4\mathrm{n_v} \mathrm{e} \mathrm{m_{eff}}} [\mathrm{u_c} \nabla \mathrm{u_c^*} - \mathrm{u_c^*} \nabla \mathrm{u_c}] \mathrm{e}^{\frac{\gamma r}{2}} \qquad (23)$$

$$u_c = e^{ik_{Li} r} + \frac{1}{k_{Li} r} e^{i(\delta + k_{Li} r)} \sin\delta + \frac{3z}{k_{Li} r^2} e^{i(\delta + k_{Li} r)} \sin\delta \qquad (24)$$

$$k_{Li} = \frac{\langle \Psi_{Li} | \hat{k} | \Psi_{Li} \rangle}{\langle \Psi_{Li} | \Psi_{Li} \rangle} \qquad k_{Si} = \frac{\langle \Psi_{Si} | \hat{k} | \Psi_{Si} \rangle}{\langle \Psi_{Si} | \Psi_{Si} \rangle} \qquad (25a,b)$$

$$\delta = -\frac{2F}{k_{Li}}\left[\frac{e^{-\alpha(r-R_e)} \sin k_{Li} r}{\alpha^2 + 4k_{Li}{}^2} (\alpha \sin k_{Li} r + 2k \cos k_{Li} r) - \frac{2k_{Li}{}^2}{\alpha^2 + 4k_{Li}{}^2} \frac{e^{-\alpha(r-R_e)}}{\alpha}\right] + \frac{F}{k_{Li}}\left[\frac{e^{-\alpha(r-R_e)} \sin k_{Li} r}{4\alpha^2 + 4k_{Li}{}^2} (2\alpha \sin k_{Li} r + 2\, k_{Li} \cos k_{Li} r) + \frac{2k_{Li}{}^2}{4\alpha^2 + 4k_{Li}{}^2} \frac{e^{-2\alpha(r-R_e)}}{2\alpha}\right] \qquad (26)$$

$$F = \frac{8 m_{eff} \pi^2 D_e}{\bar{h}^2} \qquad (27)$$

In equation 20, $E_{nm}^{Li}$ is the product of the expectation value of $x^{(m)}$, the electric charge of an electron $e$, the electric field $\vec{E}$ and scaling coefficient $S_n^{Li}$ (Sn-Li). The scaling coefficient $S_n^{Li}$ (Sn-Li) is analogous to the creation operator in quantum mechanics. Sn-Li is a function of lithium ion concentration $x$ and the energy states number in the quantum harmonic oscillator (QHO). Several scaling coefficients $S_n^{Li}$ are displayed in figure 2. The scaling coefficient $S_n^{Li}$ is linear due to the QHO when the lithiated silicon nanowire experiences approximately 20 percent volume expansion or less [6]. As the volume increases with greater lithium ion concentration $x$, Sn-Li remains linear with the energy states within the conduction bands becoming more energetic.

In equation 21, $E_{nm}^{Si}$ is the part of $E_{nm}$ that is derived from the k·p method for solving the crystallized silicon nonlinear band structures with a silicon band gap of $E_g$. The scaling coefficient $S_n^{Si}$ in equation 22 is utilized to ensure that the silicon energy component $E_{nm}^{Si}$ is comparable in magnitude and order to the lithium energy component $E_{nm}^{Li}$.

In equation 24, $u_c$ is a Bloch function that describes the lithium ion conduction band based on electron scattering theory [14]. The wave number for lithium ions $k_{Li}$ and silicon atoms $k_{Si}$ are defined in equation 25a and 25b respectively. Both wave numbers are the expectation value of the momentum operator per Planck constant for their respective lithium or silicon wave functions. In this work the average phase shift δ will be assume to be equivalent to the average scattering angle for electrons and is defined in equations 26 and 27 [15].

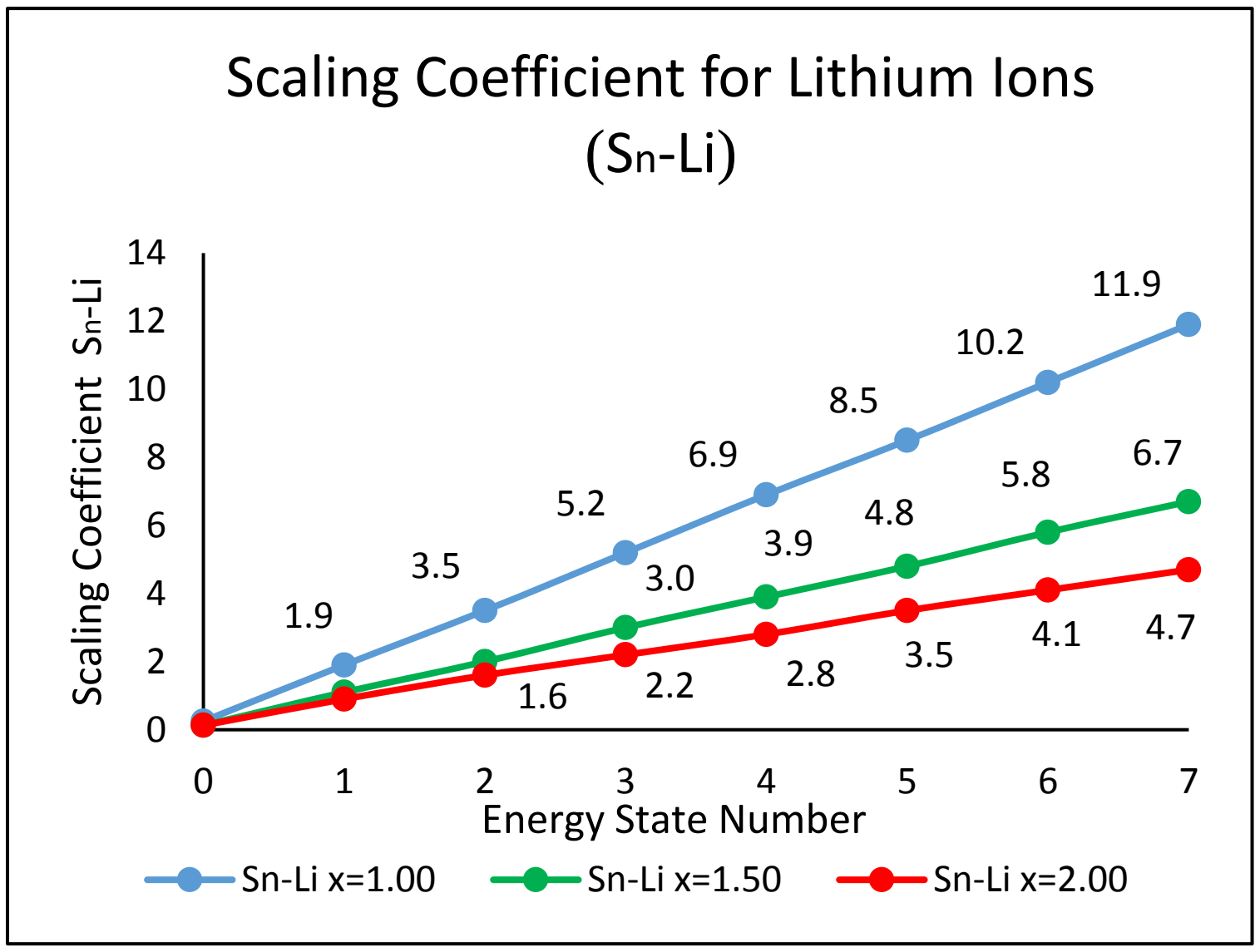

**Figure 2.** An example of three scaling coefficient (Sn-Li) which are a function of lithium concentration x. The energy state number is the label of each energy state starting with ground state 0 and energy states 1 thru 7. The scaling coefficient (Sn-Li) is dimensionless.

With the summation of equations 20 and 21, the individual energy matrix elements $E_{nm}$ within the Hamiltonian H are

$$\sum_{n=0,m=1}^{N\times M} E_{nm} = \sum_{n=0,m=1}^{N\times M} \left(E_{nm}^{Li} + E_{nm}^{Si}\right) = \mathrm{H} \qquad (28)$$

where $N$=8 are the number of energy states (ground state 0 and energy states 1 thru 7) and $M$=8 are nonlinear optical processes 1 thru 8. In order to develop the computational method for deriving second harmonic generation, the Schrödinger equation is solved as an eigenvalue equation with the Hamiltonian as the system of energy

$$\mathrm{H}\Psi_M = \lambda\Psi_N \tag{29a}$$

$$\begin{bmatrix} E_{71} & E_{72} & E_{73} & E_{74} & E_{75} & E_{76} & E_{77} & E_{78} \\ E_{61} & E_{62} & E_{63} & E_{64} & E_{65} & E_{66} & E_{67} & E_{68} \\ E_{51} & E_{52} & E_{53} & E_{54} & E_{55} & E_{56} & E_{57} & E_{58} \\ E_{41} & E_{42} & E_{43} & E_{44} & E_{45} & E_{46} & E_{47} & E_{48} \\ E_{31} & E_{32} & E_{33} & E_{34} & E_{35} & E_{36} & E_{37} & E_{38} \\ E_{21} & E_{22} & E_{23} & E_{24} & E_{25} & E_{26} & E_{27} & E_{28} \\ E_{11} & E_{12} & E_{13} & E_{14} & E_{15} & E_{16} & E_{17} & E_{18} \\ E_{01} & E_{02} & E_{03} & E_{04} & E_{05} & E_{06} & E_{07} & E_{08} \end{bmatrix} \begin{Bmatrix} \psi_1 \\ \psi_2 \\ \psi_3 \\ \psi_4 \\ \psi_5 \\ \psi_6 \\ \psi_7 \\ \psi_8 \end{Bmatrix} = \lambda \begin{Bmatrix} \psi_7 \\ \psi_6 \\ \psi_5 \\ \psi_4 \\ \psi_3 \\ \psi_2 \\ \psi_1 \\ \psi_0 \end{Bmatrix} \tag{29b}$$

**Probability Density States**

The solution to equation 29b gives the eigenenergies $\lambda_{0\text{ thru}}\,\lambda_7$ and the eigenstates $\Psi_M$ and $\Psi_N$ in which $\Psi_M = \Psi_N$. The eigenstates $\Psi_M$ and $\Psi_N$ are the summations of the nonlinear optical spin-orbitals $\psi_m$ and the energy state spin-orbitals $\psi_n$ respectively. The spin-orbitals are comprised of the spin component $\sigma_i$ which is defined as $\sigma_1 = 1 \; or \; \sigma_2 = -1$. The amplitudes $c_k$ and $c_l$ are complex numbers of $\psi_m$ and $\psi_n$ respectively.

$$\Psi_M = \sum_{m=1}^{8} \psi_m \qquad \Psi_N = \sum_{n=0}^{7} \psi_n \tag{30a, b}$$

using Einstein notation

$$\psi_m = c_k e^{-i\frac{E_{km}t}{h}} \sigma_i \quad k = 0,1,2, \dots 7 \qquad \psi_n = c_l e^{-i\frac{E_{nl}t}{h}} \sigma_i \quad l = 1,2,3, \dots .8 \tag{31a, b}$$

where k and l are summation indices. Together both eigenstates $\Psi_N$ and $\Psi_M$ define the individual probability density states $\phi_{nm}$

$$\Phi_{NM} = \langle \Psi_N | \Psi_M \rangle = \sum_{n,m}^{N\times M} \psi_n^* \psi_m = \sum_{n,m}^{N\times M} \phi_{nm} = 1 \tag{32}$$

$n$=0, 1, 2…7 and $m$=1, 2, 3...8

and equation 32 in matrix form is defined as

$$Probability\ Density\ States\ \Phi_{NM} = \begin{bmatrix} \phi_{71}\phi_{72}\phi_{73}\ \phi_{74}\ \phi_{75}\ \phi_{76}\ \phi_{77}\ \phi_{78} \\ \phi_{61}\phi_{62}\phi_{63}\ \phi_{64}\ \phi_{65}\ \phi_{66}\ \phi_{67}\ \phi_{68} \\ \phi_{51}\phi_{52}\phi_{53}\ \phi_{54}\ \phi_{55}\ \phi_{56}\ \phi_{57}\ \phi_{58} \\ \phi_{41}\phi_{42}\phi_{43}\ \phi_{44}\ \phi_{45}\ \phi_{46}\ \phi_{47}\ \phi_{48} \\ \phi_{31}\phi_{32}\phi_{33}\ \phi_{34}\ \phi_{35}\ \phi_{36}\ \phi_{37}\ \phi_{38} \\ \phi_{21}\phi_{22}\phi_{23}\ \phi_{24}\ \phi_{25}\ \phi_{26}\ \phi_{27}\ \phi_{28} \\ \phi_{11}\phi_{12}\phi_{13}\ \phi_{14}\ \phi_{15}\ \phi_{16}\ \phi_{17}\ \phi_{18} \\ \phi_{01}\phi_{02}\phi_{03}\ \phi_{04}\ \phi_{05}\ \phi_{06}\ \phi_{07}\ \phi_{08} \end{bmatrix} = 1 \qquad (33)$$

One specific computational result of the probability density states $\Phi_{NM}$ is displayed in figure 3 where the lithium ion concentration is *x*=1.00 with an applied energy Ea=2 electron-volts that serves as the energy source for the electrons that flow through lithiated silicon nanowire. For this research the focus is mainly on the second-order nonlinear optical interaction when m=2 and n=1, 2, 3…7 (red box area in figure 3). The summation of these probabilities density states $\phi_{nm}$ for the second-order nonlinear optics will be designated as $\phi_{n2}$ where n=1 through 7.

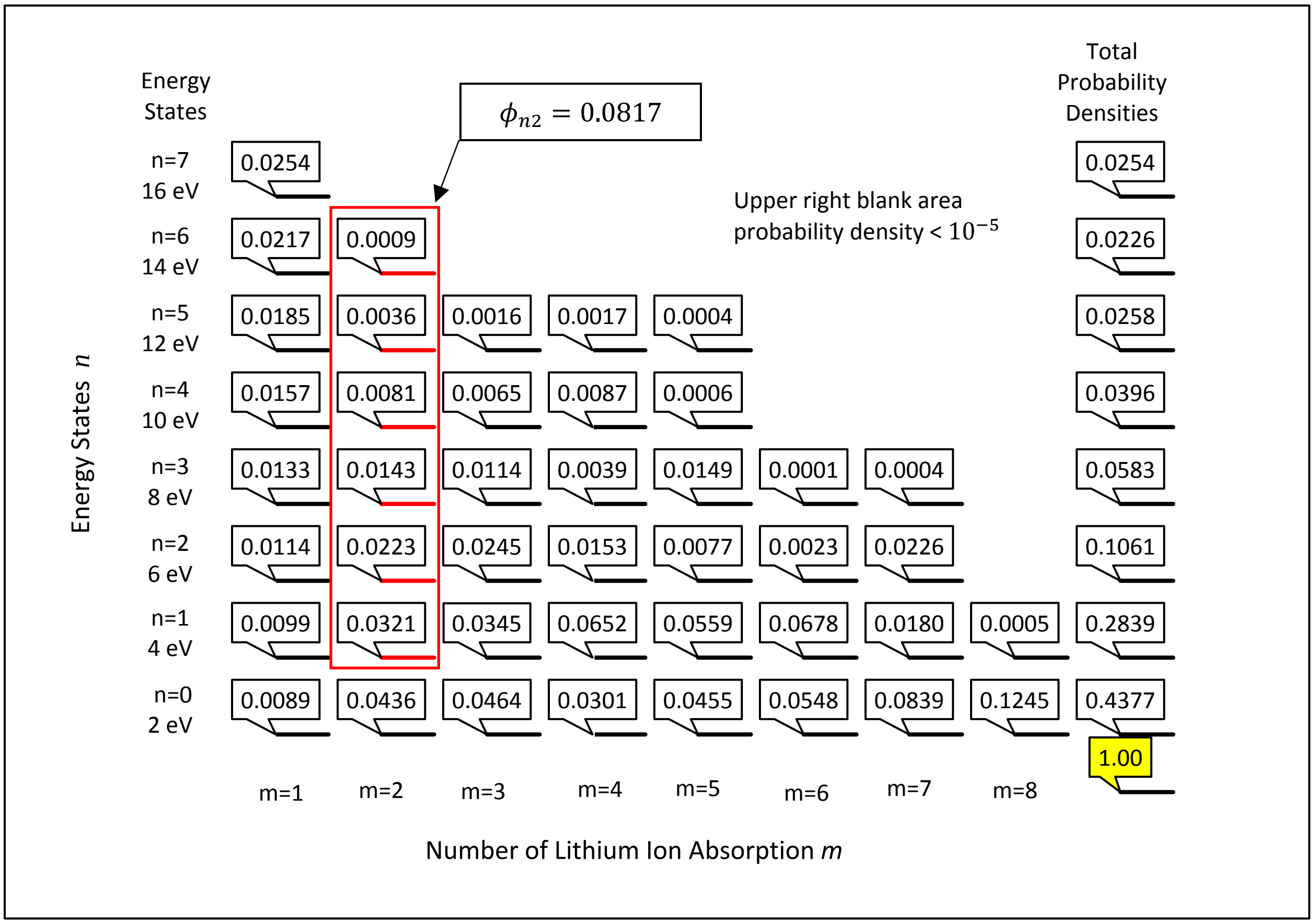

**Figure 3.** An example of one probability density states $\Phi_{NM}$with the lithium ion concentration of *x*=1.00 and applied energy of 2-eV. The total probability density $\Phi_{NM}$=1.00 which is the sum of all the entries of energies states n=0 to n=7 and number of lithium ion absorption m=1 to m=8. The sum of $\phi_{nm}$ in the red box area designated as $\phi_{n2}$, where n=1 through 7, is the probability density of second-order nonlinear optical interactions that are the combination of second harmonic generation (SHG), sum-frequency generation (SFG) and difference-frequency generation (DFG).

## Virtual Energy States

With the solution of the eigenvalue equation it is discovered that the eigenenergies $\lambda_n$ resembles a two-energy level approximation model instead of a quantum harmonic oscillator (QHO) as was intended from the construction of the Hamiltonian. As displayed in figure 4a, the QHO is formed with eight 2-eV energy states from the ground state of 2-eV to the top energy level of 16-eV. The eigenenergy $\lambda_7$ is the largest energy value and represents the top energy level for excited electron states while $\lambda_0$ is the ground state energy for electrons. The remaining eigenenergies $\lambda_n$ had values less than $10^{-17}$ eV and are therefore considered as zero or non-existent. Since there are only two eigenenergies $\lambda_n$ that are real and non-zero, the energy states $E_{nm}$ in the Hamiltonian are considered as virtual energy states. This concept fits well with the prevailing theory of nonlinear optical interactions as photon's angular frequencies are thought of as combining through the process of virtual energy states [10]. In addition, these virtual states are not considered as real since they are not known to be measurable or observable parameters [16]. Since virtual energy states are not considered as real, varies alternate probability density state configurations (figure 4b) of the two-energy level model can be produced that will yield similar results to the original QHO in figure 4a. The summation of the second-order probability density states in the QHO and the Random Virtual Energy State configuration is $\phi_{N2}$=0.125. This is due to the conservation law of probability in quantum mechanics which ensures that the summation of the probability densities in second-order *linear* optical interaction *m*=2 for each energy model of figure 4a and 4b remains constant with respect to volume and time [17].

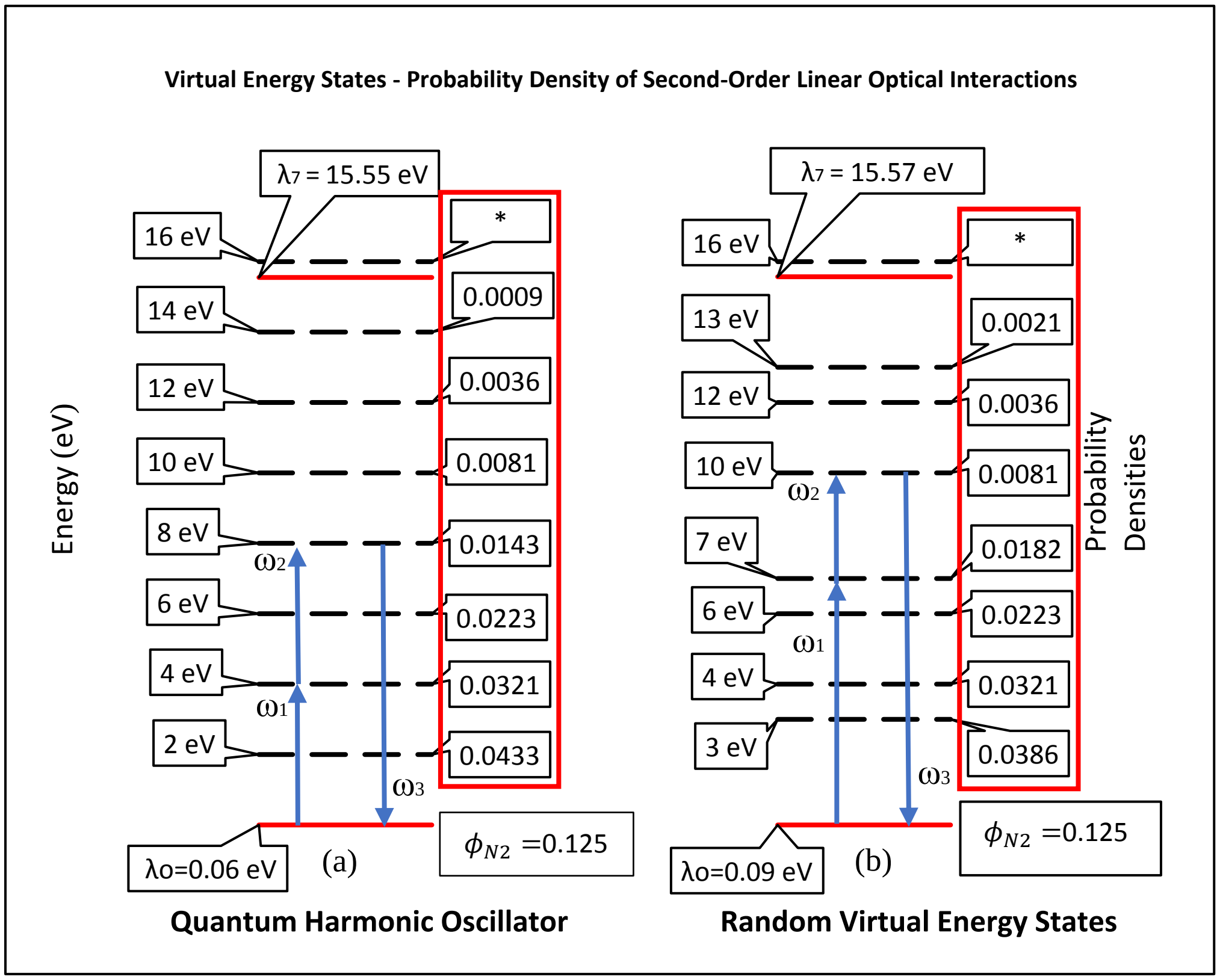

**Figure 4.** The energy matrix elements $E_{nm}$ in equation 29b are virtual energy states as displayed by the black dash lines. The red solid lines are the real energy states of a two energy level approximation model that are eigenvalue energies. The summation of both probability densities of second-order *linear* optical interactions $\phi_{N2}$ in (a) quantum harmonic oscillator and (b) random virtual energy states model are 0.125 (red box area) due to the law of conservation of probability. The two examples of second-order nonlinear optical interactions are $\omega_3 = \omega_2 + \omega_1$ in (a) second harmonic generation (SHG) where $\omega_2 = \omega_1$ and in (b) sum-frequency generation (SFG) where $\omega_2 \neq \omega_1$. The asterisk * is for probability density states less than $10^{-5}$.

The probability density of the second-order *nonlinear* optical interaction $\phi_{n2}$ as shown in figure 3 is a subgroup of the probability density of second-order *linear* optical interactions $\phi_{N2}$ as seen in figure 4. The probability density of the ground state is not included in $\phi_{n2}$ [18]. When the applied energy source Ea increases from Ea=2-eV to Ea=4-eV, in order to calculate the probability density states $\Phi_{NM}$ the Hamiltonian matrix NxM will change from 8x8 to 4x4 as the upper energy level remains at 16-eV. The result is a slight increase in $\phi_{n2}$ between the two energy sources. However for both applied energy sources Ea, it is discovered that with increasing lithium ion concentration $x$ that $\phi_{n2}$ decreases slightly. It can be concluded that the increase in applied energy Ea and lithium ion concentration $x$ that the probability density of second-order nonlinear optical interaction $\phi_{n2}$ approaches zero. In figure 5 the summation of $\phi_{n2}$ for both applied energies of 2-eV and 4-eV and for the lithium ion concentrations of x=1.00, 1.50 and 2.00 are displayed.

| | | $\phi_{n2}$ | | |
|---|---|---|---|---|
| Ea=2-eV | | x=1.00 | x=1.50 | x=2.00 |
| Energy States | $E_n$ | NxM 8x8 | NxM 8x8 | NxM 8x8 |
| Virtual State 7 | 16 | * | * | * |
| Virtual State 6 | 14 | 0.0009 | 0.0009 | 0.0008 |
| Virtual State 5 | 12 | 0.0036 | 0.0034 | 0.0032 |
| Virtual State 4 | 10 | 0.0081 | 0.0080 | 0.0080 |
| Virtual State 3 | 8 | 0.0143 | 0.0142 | 0.0138 |
| Virtual State 2 | 6 | 0.0223 | 0.0216 | 0.0198 |
| Virtual State 1 | 4 | 0.0321 | 0.0330 | 0.0336 |
| **Second-Order Nonlinear Optical Probability Density** | $\phi_{n2}$ | **0.0817** | **0.0811** | **0.0792** |

| | | $\phi_{n2}$ | | |
|---|---|---|---|---|
| Ea=4-eV | | x=1.00 | x=1.50 | x=2.00 |
| Energy States | $E_n$ | NxM 4x4 | NxM 4x4 | NxM 4x4 |
| Virtual State 3 | 16 | 0.0010 | 0.0008 | * |
| Virtual State 2 | 12 | 0.0186 | 0.0172 | 0.0166 |
| Virtual State 1 | 8 | 0.0712 | 0.0706 | 0.0698 |
| **Second-Order Nonlinear Optical Probability Density** | $\phi_{n2}$ | **0.0908** | **0.0886** | **0.0864** |

**Figure 5.** The probability density of second-order nonlinear optical interaction $\phi_{n2}$ is shown for varies applied energies Ea and lithium ion concentrations $x$. For applied energy source 4-eV the Hamiltonian is 4x4 which results in a marginal increase in the probability density compare to Ea=2-eV which represents a Hamiltonian that is a 8x8 matrix. The asterisk * is for probability density states less than $10^{-5}$.

**Second Harmonic Generation**

As mentioned previously, second-order nonlinear optical interaction is a combination of three processes: second harmonic generation (SHG), sum-frequency generation (SFG) and difference-frequency generation (DFG). In order to distinguish between the three nonlinear optical processes a set of equations has been derived from the quantum mechanical principle that photons are indistinguishable particles [19].

$$\alpha_{SHG} = \frac{1}{2} \qquad \alpha_{SFG} = \frac{1}{4} + \frac{\omega_2}{4\omega_1} \qquad \alpha_{DFG} = \frac{1}{4} - \frac{\omega_2}{4\omega_1} \qquad (34\text{ a, b, c})$$

where $\omega_1 \geq \omega_2$ and $0 \leq \frac{\omega_2}{4\omega_1} \leq \frac{1}{4}$ $\qquad \alpha_{SHG} + \alpha_{SFG} + \alpha_{DFG} = 1$

$$\phi_{SHG} = \alpha_{SHG} * \phi_{n2} \qquad \phi_{SFG} = \alpha_{SFG} * \phi_{n2} \qquad \phi_{DFG} = \alpha_{DFG} * \phi_{n2} \qquad (35\text{ a, b, c})$$

In equation 34, a series of coefficients has been derived based on this quantum mechanical principle that are a function of the second-order nonlinear optical angular frequencies $\omega_1$ *and* $\omega_2$. The coefficient of $\alpha_{SFG}$ and $\alpha_{DFG}$ is dependent on these angular frequencies however, the second harmonic generation coefficient $\alpha_{SHG}$ is actually constant 1/2 as seen in figure 34a. Therefore probability density of SHG $\phi_{SHG}$ is one half the probability density of the second-order of nonlinear optical interaction $\phi_{n2}$ as seen in equation 35a.

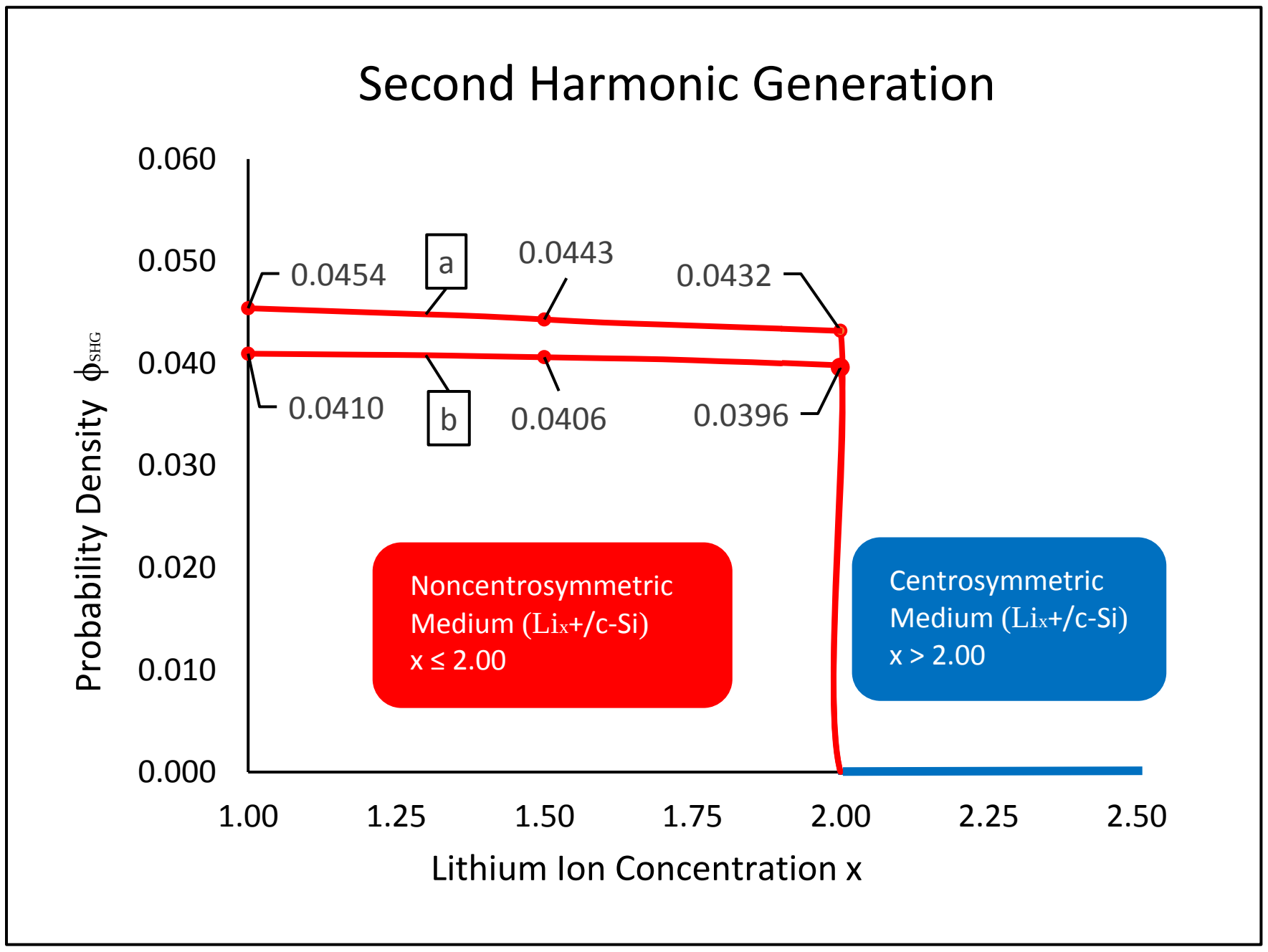

**Figure 6** The curves (a) and (b) describes the probability densities of SHG with an applied energy of Ea=4-eV and Ea=2-eV respectively. The lithiated silicon material is considered to be non-centrosymmetric when the lithium ion concentration is $x \leq 2.00$ which can support second harmonic generation. When $x > 2.00$ the lithiated silicon material transforms to a centrosymmetric medium at which point SHG does not exist and probability density of $\phi_{SHG}$ is zero.

As previously stated, the non-centrosymmetric medium in which second harmonic generation (SHG) does exist, lithium ions diffuses through crystallized silicon ($Li_x$+/c-Si) with lithium concentration *x* defined as the ratio of lithium ions to crystallized silicon atoms. Once the Li+ chemical reacts with c-Si to form amorphous lithiated silicon (a-$Li_x$Si) this material becomes inversion symmetric and transform into centrosymmetric medium in which SHG does not exist as shown previously in figure 1. In figure 6, theoretically starting at *x*=1.00, as the lithium ion concentration *x* increases in the $Li_x$+/c-Si medium the SHG will approach zero if the medium remains as a non-centrosymmetric material. However, as *x* reaches 2.00 the SHG will suddenly decrease to zero as the $Li_x$+/c-Si medium changes to centrosymmetric material.

## Summary

The preceding computational method in this research has been demonstrated to calculate the probability density states of second harmonic generation in lithiated silicon nanowires. The possibility for SHG to manifest is low with approximately 0.0396 to 0.0454 probability per unit volume of occurring. Nevertheless, this may make it a candidate for the application of second harmonic imaging microscopy with the proper filter in order to isolate this second-order nonlinear optical process [20].